\documentclass{aa}
\usepackage{graphicx}
\begin{document}
	\title{Search for the elusive optical counterpart of PSR J0537--6910
with the HST Advanced Camera for Surveys}

   \author{
	R. P. Mignani\inst{1}
	\and
          L. Pulone\inst{2}  
	\and
	G. Iannicola \inst{2}
	  \and
       G. G. Pavlov\inst{3}
	 \and
       L. Townsley\inst{3}
	 \and
       O. Y. Kargaltsev\inst{3}
}

   \offprints{R. P. Mignani}

   \institute{
	European Southern Observatory, Karl-Schwarzschild-Str.\,2, 
	D-85748, Garching, Germany \\
              \email{rmignani@eso.org}          
                  \and
            INAF-Osservatorio Astronomico di Roma, Via di Frascati 33,   
I-00040 Monte Porzio Catone, Italy, \\ 
\email{pulone@mporzio.astro.it,giacinto@mporzio.astro.it}
  \and
	 Pennsylvania State University, 525 Davey Lab, University Park, PA 16802, USA,\\
\email{pavlov@astro.psu.edu, townsley@astro.psu.edu, 
green@astro.psu.edu} 
}

   \date{Received ...}

\titlerunning{PSR J0537-6910}

   \abstract{We   present  the   results  of   deep,  high-resolution,
multi-band  optical observations  of  the field  of  the young  ($\sim
5,000$ yrs)  16 ms  X-ray pulsar PSR  J0537$-$6910 performed  with the
Advanced  Camera  for  Surveys  (ACS)  aboard the  {\sl  Hubble  Space
Telescope}  ({\sl HST}).   Although a  few new  potential counterparts
have been detected within or  close to the revised {\sl Chandra} X-ray
error  circle ($\simeq 1  \arcsec$) of  the pulsar,  only two  of them
(with magnitudes $m_{814W} \approx  23.9$ and $m_{814W} \approx 24.2$)
show  indications of  a peculiar  spectrum which  could be  related to
optical emission from the pulsar. This might be true also for a third,
fainter,  candidate  detected  only  in  one  filter  (with  magnitude
$m_{814W} \approx 26.7$).  If either of the two brighter candidates is
indeed the actual counterpart,  the optical output of PSR J0537$-$6910
would make  it similar  to young Crab-like  pulsars. If not,  it would
mean that PSR J0537$-$6910 is significantly underluminous with respect
to  all pulsars  detected in  the optical.   \keywords{Stars: pulsars:
individual: PSR J0537$-$6910 } }

   \maketitle
%

\section{Introduction}

PSR~J0537$-$6910  is  a  young  X-ray  pulsar at  the  center  of  the
plerionic supernova remnant N157B in  the LMC, close to the 30 Doradus
star   forming  region.   Pulsations   with  a   16  ms   period  were
serendipitously discovered  by {\sl RXTE} (Marshall et  al.\ 1998) and
immediately confirmed  by {\sl ASCA} and {\sl  Beppo-SAX} (Cusumano et
al.\  1998). Thus,  PSR~J0537$-$6910 has  taken over  the  Crab pulsar
($P=33$ ms) as  the fastest ``ordinary'' pulsar (i.e.,  not spun up by
accretion  from  a  companion  star)  discovered  so  far.   The  time
derivative   of  the  pulsar   period  ($\dot   P  \approx   5  \times
10^{-14}$s~s$^{-1}$), obtained  from the multi-epoch  timing (Marshall
et al.\ 1998; Cusumano et al.\ 1998) yielded a spin down age of $ \sim
5\,000$~yrs, similar  to the  age of N157B  (Wang and  Gotthelf 1998),
which  makes PSR~J0537$-$6910  one  of the  handful  of known  pulsars
younger than 10\,000 yrs.  At the same time, with a derived rotational
energy loss $\dot  E \approx 4.8 \times 10^{38}$  ergs~s$^{-1}$, it is
the  most energetic  pulsar known.   As expected  for a  young pulsar,
large glitches  in the spindown  have been repeatedly  detected during
different  {\sl  RXTE} timing  campaigns  (Middleditch  et al.\  2001;
Marshall  et al.\ 2004).   Observations of  PSR~J0537$-$6910 performed
with the {\sl Chandra X-ray  Observatory} (Wang et al.\ 2001; Townsley
et al.\  2004) have resolved  a compact ($\simeq  2\farcs5\times 7''$)
pulsar-wind nebula (PWN) elongated  perpendicular to the symmetry axis
of  a  larger comet-like  structure  ($\simeq  40\arcsec$).  This
structure  can be  explained as  a trail  left behind  by  the pulsar
during its motion in the  ISM.  Although its large energy output makes
PSR~J0537$-$6910 a natural target for multiwavelength observations, it
has not  yet been detected outside  the X-ray band.   So far, searches
for its radio counterpart yielded null results, with an upper limit of
$F_{1.4\,  {\rm GHz}} \sim  0.04$~mJy (Crawford  et al.\  1998), which
suggests that its radio  luminosity is significantly weaker than those
of  the  Crab pulsar  and  of the  other  young  LMC pulsar  B0540-69.
Searches for an optical counterpart of PSR~J0537$-$6910 have also been
unsuccessful so  far.  First observations  performed with the  ESO NTT
(Mignani et al.\  2000) and the ESO 3.6m  (Gouiffes \& \"Ogelman 2000)
telescopes failed  to detect  any potential pulsar  counterpart within
the original  {\sl ROSAT} $5''$  error circle (Wang \&  Gotthelf 1998)
down to $V  \sim 23.4$.  The same conclusion was  reached by Butler et
al.\ (2002)  using archived  {\sl HST} observations  of the  field and
exploiting  the updated {\sl  Chandra} pulsar  position (Wang  et al.\
2001).  Here  we report  on the results  of recent, much  deeper, {\sl
HST} observations of the PSR J0537$-$6910 field.  The same dataset has
been  also used to  search for  the optical  counterpart of  the X-ray
Pulsar Wind Nebula (PWN); the results will be reported elsewhere.  The
observations  and  data  analysis   are  described  in  \S2  and  \S3,
respectively, while the results are  presented in \S4 and discussed in
\S5.

\section{Observations}

The field  of PSR J0537$-$6910 was  observed on January  2003 with the
Wide Field Channel  (WFC) of the Advanced Camera  for Surveys (ACS) on
board the {\sl HST}.  The  ACS/WFC, a two-CCD $4096 \times 2048$ pixel
detector (Pavlovsky  et al.\  2003), has a  combined field of  view of
$3\farcm 3\times 3\farcm 3$ and an average pixel size of $0\farcs050$.
Observations were  performed during  two consecutive {\sl  HST} orbits
through  the  435W  ($\lambda=4297$\,\AA;  $\Delta\lambda=1038$\,\AA),
555W   ($\lambda=5346$\,\AA;   $\Delta\lambda=1193$\,\AA)   and   814W
($\lambda=8333$\,\AA; $\Delta\lambda=2511$\,\AA)  filters, selected to
maximize the  spectral coverage.  For  each filter, sequences  of four
exposures were  taken to  allow cosmic ray  filtering and  to minimize
saturation effects of bright stars  in the field close to the position
of our target.  The corresponding total integration times were 3200 s,
2800   s  and   2800  s   in  the   435W,  555W   and   814W  filters,
respectively. Standard  data reduction (debiassing,  flatfielding) and
photometric  calibration were  applied according  to the  standard ACS
pipeline.  The  default ACS  photometric calibration available  at the
time of  image acquisition was later  updated by using  as a reference
the  new instrument  zero-points computed  from the  improved Detector
Quantum Efficiency  (DQE) and filter  throughput curves (De  Marchi et
al.\ 2004).  For  each filter, the STSDAS {\it  drizzle} task was used
to combine  single exposures and produce  a mosaic image  from the two
ACS/WFC chips  after correcting for  the geometric distortions  of the
CCDs.

\section{Data analysis}

\subsection{HST ACS Astrometry}

The absolute uncertainty of the  astrometric solution in the {\sl HST}
focal plane is  still based on guide stars from  the GSC1.1 (Lasker et
al.\ 1990) and  is of the order of 1\farcs0  (Pavlovsky et al.\ 2003).
For this reason,  we reassessed the ACS astrometry  using more precise
catalogs        as       a       reference.         Although       the
GSC2\footnote{http://www-gsss.stsci.edu/gsc/gsc2/GSC2home.htm}    seems
to be the first natural choice for this purpose, it has been discarded
because of coarser astrometric  accuracy in this specific field.  This
is  likely caused  by systematics  affecting  the catalog  in the  LMC
region, related to both the local crowding and the presence of diffuse
nebulosities  associated with  the  N157B supernova  remnant.  On  the
other  hand, the recently  released USNO-B1.0  catalog (Monet  et al.\
2003) seems  also to  be affected  by  some kinds  of systematics  and
provides only a  poor and uneven coverage of the  region, with most of
the reference  stars located  at the  edge of the  ACS field  of view.
Therefore, we  decided to  use the Two  Micron All Sky  Survey (2MASS)
Point Source catalog (Skrutskie et  al.\ 1997).  After matching on the
ACS images the sky and  pixel coordinates of 30 well distributed 2MASS
stars,  a revised  astrometric solution  has been  computed  using the
package
ASTROM\footnote{http://star-www.rl.ac.uk/Software/software.htm},
yielding an overall accuracy of $\sim 0\farcs1$ per coordinate.  As an
independent check, we qualitatively evaluated the astrometric solution
using as  a reference  the positions of  stars selected from  the very
accurate ($0\farcs035$)  UCAC2 catalog (Zacharias et  al.\ 2004).  The
matches between  the actual  and expected positions  were found  to be
consistent within the uncertainty  of our astrometric calibration.  We
note that since the UCAC2 is a relatively bright catalog ($V \le 16$),
all the  matched stars are saturated  in the ACS images  and cannot be
used directly for astrometric calibration.

\subsection{Chandra ACIS Astrometry}

\begin{figure}[h]
   \centering
      \includegraphics[bb=10 200 460 660,width=8cm,clip]{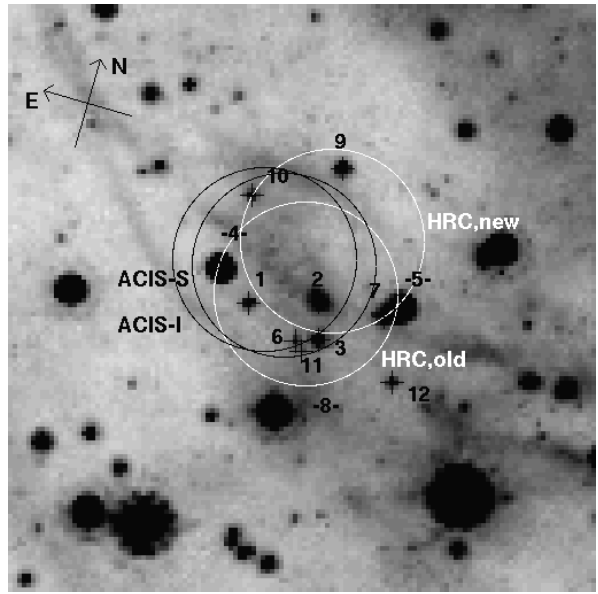}
     \caption{Field  around PSR  J0537$-$6910  ($6\farcs3 \times 6\farcs3$) imaged  with the  ACS
filter 814W  (2800 s exposure).   The four  circles indicate the X-ray
positions  of the pulsar (see   Table 1) as  derived from observations
performed  with the {\sl Chandra} HRC  (white; `old' from Wang et al.\
2001;   `new' from Gotthelf   2003,  private  communication) and  ACIS
(black;   Townsley  et  al.\  2004).  For    each   circle, the radius
corresponds  to the  overall    uncertainty  of the pulsar    position
resulting from  the  combination  of the   estimated error   of  the {\sl
Chandra} coordinates ($\sim   1''$)  and the ACS  astrometric  solution
($0\farcs1$).  Stars 4, 5 and 8 (marked by horizontal ticks) are stars
6, 3 and 4 of Mignani et al.\ (2000). } 
         \label{finding-chart} 

\end{figure}

As a reference for the search of the  pulsar counterpart we 
use  the most  precise  X-ray positions  obtained  with {\sl  Chandra}
(Weisskopf  et  al.\  2002).   The  previous  {\sl  Chandra}  position
obtained with the imaging array  of the High Resolution Camera (HRC-I)
by Wang et  al.\ (2001) has been recently  reassessed by improving the
aspect solution  (Gotthelf 2003, private  communication).  Independent
position determinations  have also been obtained  through the analysis
of  ACIS (Advanced  CCD Imaging  Spectrometer; Garmire  et  al.\ 2003)
observations,  taken   in  the  ACIS-I  (imaging   array)  and  ACIS-S
(spectroscopy  array) modes  (Townsley  et al.\  2004).  The  original
ACIS-I data (ObsID  62520), taken in September 1999,  were centered on
the  R136  stellar  cluster  in  30 Doradus.   The  observations  were
performed  in  ``alternating'' mode  consisting  of  sequences of  ten
3.3~sec frames followed by a single 0.3~sec frame to eliminate pile-up
in  N157B.   We  used  the  short frames  of  the  ACIS-I  observation
integrated  over the  full energy  band (0.5--8~keV)  to  estimate the
pulsar position.  Unfortunately, the  small number of counts available
($\approx 20$)  prevented us from obtaining the  position in hard-band
channels, so that counts from the  PWN and/or the SNR may be affecting
the  determination  of the  actual  pulsar  centroid.   To reduce  the
uncertainty of  the pulsar  position, we tried  to improve  the ACIS-I
absolute   astrometry  by   cross-matching  the   coordinates   of  30
serendipitous   X-ray   sources   with   the   2MASS   catalog,   but,
unfortunately,  found only  4 high-confidence  matches, which  did not
allow  us to  obtain a  boresight correction  more precise  than $\sim
0\farcs3$.    Systematic uncertainties associated   with the  relatively large
off-axis angle may  also affect  the  absolute position of the  pulsar
measured from the ACIS-I data.  Such systematics have been measured in
the  Chandra Deep Field North   (Alexander et al.   2002) and in Orion
(Getman et al.  2005) and at an off-axis angle of $5\arcmin$ they turn
out to be    $\approx  0\farcs2$.

Given    these   caveats,   the    estimated   pulsar    position   is
$\alpha(J2000)=05^{\rm  h}37^{\rm  m}47\fs46$, $\delta(J2000)=-69\degr
10\arcmin 19\farcs9$, with an  approximate error of $1\farcs0$, which
accounts  for  both  the  errors   of  the  pulsar  centroid  and  the
uncertainty  of  our  boresight  correction.  The  ACIS-S  observation
(ObsID  2783) had  N157B at  the aim  point, and  it was  taken  in a
subarray  mode to  minimize  pile-up.  The  pulsar centroid,  measured
using {\em wavdetect} over the full energy band (0.5--8~keV), yields a
position    of   $\alpha(J2000)=    05^{\rm    h}37^{\rm   m}47\fs42$,
$\delta(J2000)=  -69\degr 10\arcmin  20\farcs2$.  The  same procedure,
but  cutting in the  3--7~keV energy  range to  minimize contamination
from the PWN/SNR background,   yields a position of $\alpha(J2000)=
05^{\rm   h}37^{\rm   m}47\fs39$,   $\delta(J2000)= -69\degr
10\arcmin  20\farcs1$, i.e. consistent with the previous one.
The errors on  these positions lack the systematics present
in the off-axis ACIS-I observation.  However, due to the smaller field
of  view in sub-array  mode, too  few X-ray  sources were  detected to
compute the  boresight correction by  cross-matching with 2MASS  or to
register the  ACIS-S frame onto the  ACIS-I frame.  As  the effects on
the absolute astrometry  are roughly of the same  magnitude, we assume
that  the global errors  on the  ACIS-S positions  are similar  to the
ACIS-I one (about $1\farcs0$).

\begin{table}
\caption[]{Chandra coordinates of PSR J0537$-$6910. In all cases an
error radius of $1"$ is estimated (see text).}
\begin{center}
\begin{tabular}{llll} \\ \hline 
Source  & $\alpha(J2000)$ & $\delta(J2000)$ & References  \\ 
        & ($^{\rm  h} ~ ^{\rm  m} ~ ^{\rm  s}$) & ($\degr ~ \arcmin ~ \arcsec)$  & \\ \hline 
HRC-I   & 05 37 47.36 & $-$69 10 20.4  &  Wang et al.\ (2001) \\
HRC-I   & 05 37 47.34 & $-$69 10 19.8  &  Gotthelf (p.c.) \\
ACIS-I  & 05 37 47.46 & $-$69 10 19.9  &  Townsley et al.\ (2004)\\
ACIS-S  & 05 37 47.42 & $-$69 10 20.2  &  present work  \\ \hline
\end{tabular}
\end{center}

\end{table}

 The available estimates  for   the  {\sl Chandra}  position   of  PSR
J0537$-$6910  (Table 1)   are    all consistent within  the   computed
errors. Figure 1 shows the HRC  and ACIS pulsar positions superimposed
on the ACS/WFC 814W image  after astrometric recalibration.  A few objects
(numbered 1 through 12) can be identified within or close to the error
circles.  The three candidates already investigated by Mignani et al.\
(2000) are highlighted  in Figure 1  (see caption).  Because of  their
faintness and proximity    to brighter stars,  all  the  other objects
identified within the error circles could not  be detected in the less
deep and   more confused ground-based  NTT  observations by Mignani et
al.\ (2000).

\begin{figure*}[t]
   \centering
      \includegraphics[bb=10 150 600 720,width=8.8cm,clip]{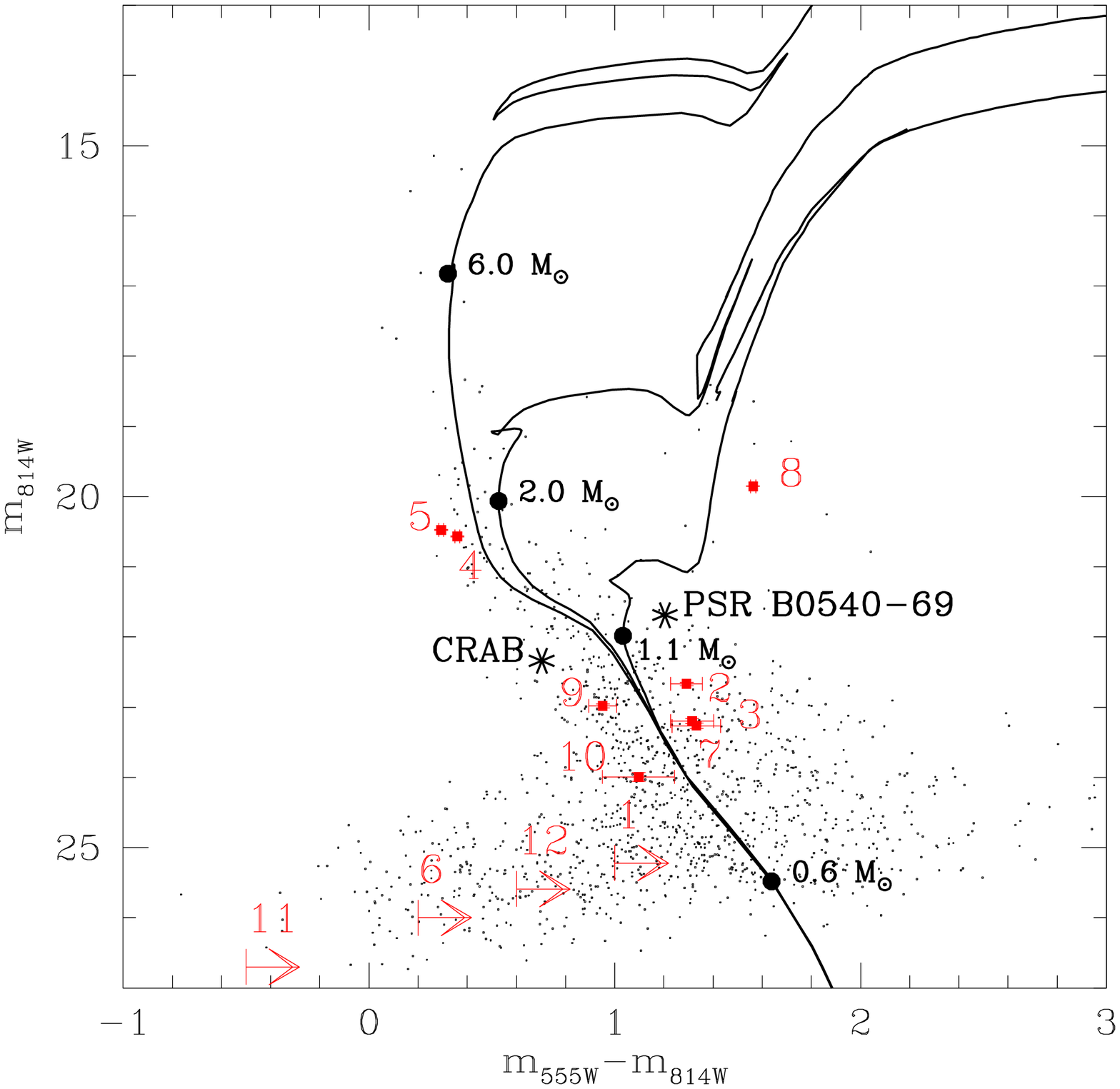} 
      \includegraphics[bb=10 150 600 720,width=8.8cm,clip]{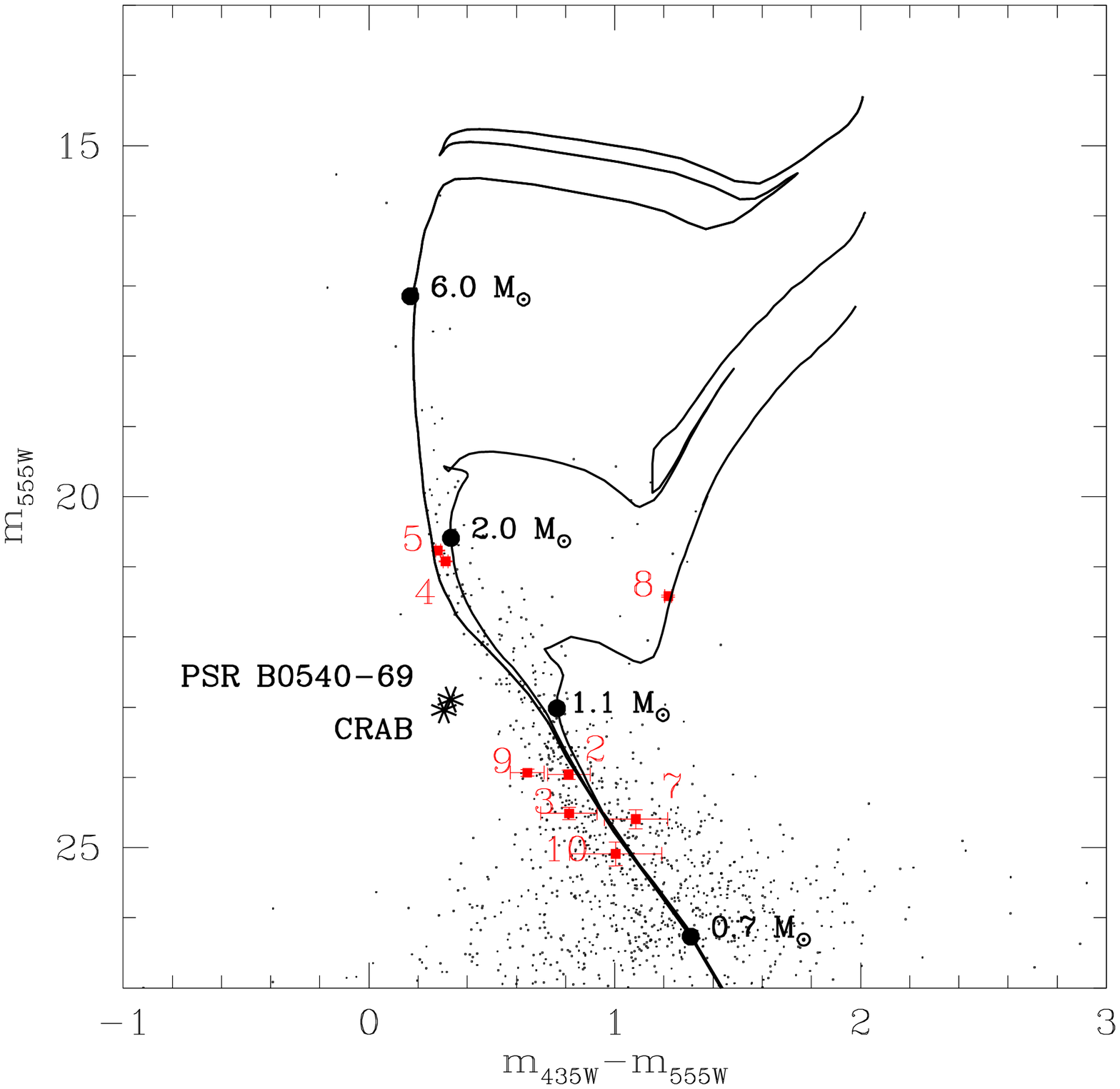}
     \caption{
Observed Color-Magnitude Diagrams (CMDs) of all the stars
detected at $\ge 5 \sigma$ significance within a $40'' \times 40''$ 
box centered on the pulsar position. 
  Field stars are marked by small dots.
Objects 2 through 5 and 7 through 10 (Figure 1) are marked by
filled squares and numbered accordingly.  In the left diagram, the arrows indicate the 
limits on the locations of objects 1, 6, 11 and 12. 
As a reference, the locations of the Crab and PSR B0540$-$69 are also plotted (asterisks)
after renormalizing the fluxes to the N157B distance and extinction. The lines indicate stellar isochrones  
computed from theoretical models
(Salasnich et al.\ 2000) for different age values 
(starting from the top: 60 Myr,  600 Myr, and 3 Gyr, respectively)
and reddened for an assumed $E(B-V)$=0.32 (see text). The filled circles indicate the locations
      of stars at the Turn-off (1.1, 2 and 6 $M_{\odot}$) and of stars close to the detection limit in the 814W and 555W passbands (0.6 and 0.7 $M_{\odot}$, respectively).}
         \label{spectra} 
\end{figure*}

\subsection{ACS Photometry}

To perform  accurate photometry  on the ACS  images, we run  an object
detection program using the  ROMAFOT package for photometry in crowded
fields  (Buonanno \&  Iannicola  1989).  The  ROMAFOT parameters  were
tuned to  achieve a $\ge 5\sigma$  object detection in  each filter. A
template PSF was obtained by fitting the intensity profiles of some of
the brightest, unsaturated, isolated stars  in the field with a Moffat
function, plus  a numerical  map of the  residual to better  take into
account the contribution of the  PSF wings.  As a reference for object
detection, we used  our 814W-band image, for which  the effects of the
local absorption are  minimized.  The master list of  objects was then
registered on the  images taken in the 435W and  555W filters and used
as a reference for the fitting  procedure.  A careful check by eye was
performed to  ensure that  all the stellar  objects found in  the 814W
band were  successfully fitted in the  other images and  to filter out
spurious  detections.  Aperture corrections  were computed  by fitting
the growth  curve of a  suitable set of standard  stars.  Instrumental
magnitudes       were        then       converted       into       the
STMAG\footnote{http://www.stsci.edu/hst/acs/analysis/zeropoints/\#zeropoints}
photometric system by applying the updated ACS zero points given by De
Marchi et al.\ (2004).

\section{Results}

The observed  magnitudes and colors of all  the potential counterparts
identified in  Figure 1 are summarized  in Table 2.   No other objects
were identified  within or  close to the  {\sl Chandra}  error circles
down to $5  \sigma$ limiting magnitudes of about  26.4, 26.2, and 27.2
in the 435W,  555W, and 814W passbands, respectively.  \\ We note that
the photometry of  star 7 is likely affected  by an uncertainty larger
than the  attached formal errors as  it is partially  blended with the
brighter star  5. The same  is probably true  also for object  2 which
might actually be a doublet of unresolved objects, as suggested by its
slightly elongated  shape.  However, our  PSF analysis is  affected by
the presence  of a  relatively bright filament  crossing the  object 2
position.  Objects 1, 6, 11 and  12 have been clearly detected only in
the  814W passband and  are probably  very reddened.  The interstellar
extinction toward PSR J0537$-$6910 is  relatively uncertain.Taking as
a  reference the  measurements of  the Balmer  decrement  around N157B
(Caplan \& Deharveng 1985), we can derive a color index $E(B-V)$=0.32.
On the  other hand, from the X-ray  spectral fits to the  ACIS data of
PSR J0537$-$6910 we  derive a hydrogen column density  $N_{\rm H} \sim
(0.5$--$1.0)\times 10^{22}$ cm$^{-2}$, depending on the assumptions on
the  metal  abundance  (see  \S5).   By adopting  the  extinction  law
appropriate for the 30 Doradus region (Fitzpatrick 1986), these values
correspond  to $E(B-V)=0.22$--0.42.   This gives  an  average $E(B-V)$
consistent  with that derived  from Caplan  \& Deharveng  (1985).  For
this reason, we will use $E(B-V) = 0.32$ as an indicative value,
 with an uncertainty of $\pm$ 0.1,
 
To assess  the  nature  of  the
candidates listed in  Table 2, we compare their  locations in the 814W
vs.\  555W$-$814W and 555W  vs.\ 435W$-$555W  Color-Magnitude Diagrams
(CMDs)  with those  derived from  the photometry  of  a representative
number of  field stars.   Figure 2  shows the CMDs  built for  all the
stars  detected at  $\ge  5 \sigma$  in  a $40''  \times 40''$  region
centered on the  pulsar position, including objects 2  through 5 and 7
through 10  which were detected in  all the passbands  (Table 2).  For
convenience, the  computed magnitudes  and colors are  renormalized to
the  VEGAMAG$^{3}$ photometric  system  because it  is  closer to  the
Johnson-Cousins system.   The conversion from the STMAG  to VEGAMAG is
done  by  applying  zero-point  offsets  of  0.61,  0.04,  and  $-$1.29
magnitudes  for the  435W, 555W,  and 814W  filters,  respectively (De
Marchi et  al.\ 2004).  As a  reference, we overplot a  set of stellar
isochrones computed  from theoretical models of metallicity $Z$=0.008
and helium abundance Y=0.23 (Salasnich  et al.\ 2000)
for different  age values (60  Myr, 600 Myr  and 3 Gyr).   The assumed
stellar masses  range from  $6 M_{\odot}$ for  the stars close  to the
Turn-off of  the $60$ Myr isochrone,  to $\sim 0.6  M_{\odot}$ for the
faintest  detected objects.   For a  direct comparison
with the observed points,  an interstellar extinction corresponding to
the assumed $E(B-V)$=0.32 has been applied to the isochrones.
Most of  the objects detected close  to the pulsar  position lie along
the Main  Sequence (MS).  The  only exception is  object 8 (star  4 of
Mignani et al.\ 2000) that lies on the giant branch of the $3$\,Gyr isochrone.

Thus, none of these
objects is  distinguishable from the local LMC  stellar population and
can  be identified  as  a  pulsar counterpart  based  on the  observed
colors.  Unfortunately, very little can  be said about the  nature of
objects 1, 6, 11 and 12.  Based on the 
555W$-$814W  
lower limits (Fig.\ 2),
only object 11  (and perhaps 6) might be  substantially bluer compared
to the MS  and field stars, but this can only  be verified with deeper
observations.

\begin{table}
\begin{center}
\begin{tabular}{lccc} \\ \hline 
ID  & $m_{\rm 814W}$ & $m_{\rm 555W}-m_{\rm 814W}$ & $ m_{\rm 435W} - m_{\rm 555W}$   \\ \hline
   1  &    25.22     (0.04)  &    $\ge +1.0$         &     -              \\
   2  &    23.96     (0.02)  &    $-$0.04    (0.07)  &     +0.24    (0.09)\\
   3  &    24.49     (0.03)  &    $-$0.01    (0.09)  &     +0.24    (0.12)\\
   4  &    21.85     (0.01)  &    $-$0.97    (0.01)  &    $-$0.26    (0.01)\\
   5  &    21.77     (0.01)  &    $-$1.03    (0.01)  &    $-$0.29    (0.01)\\
   6  &    26.00     (0.11)  &    $\ge +0.2$          &    -\\
   7  &    24.55     (0.04)  &    0.00       (0.10)  &     +0.51    (0.17)\\
   8  &    21.14     (0.01)  &    +0.23      (0.01)  &     +0.64    (0.02)\\
   9  &    24.27     (0.03)  &    $-$0.38    (0.06)  &     +0.07    (0.07)\\
  10  &    25.28     (0.06)  &    $-$0.23    (0.15)  &     +0.43    (0.22)\\
  11  &    26.70     (0.21)  &    $\ge -0.5$         &     -\\ 
  12  &    25.59     (0.08)  &    $\ge +0.6$          &     -  \\ \hline    
\end{tabular}
\end{center}
\caption[]{Observed
magnitudes, colors and associated errors
(in parentheses) for all the objects
 identified in Figure 1 which have been detected at a $\ge 5 \sigma$ level
 in at least one passband. Magnitudes and colors 
are computed in the STMAG$^{3}$ photometric system.}
\end{table}

To assess the spectral  energy distributions (SEDs) of the candidates,
we computed  their extinction-corrected  spectral fluxes at  the central
wavelengths  of  the ACS  filters.   They  are  plotted in  Figure  3,
together with the $5\sigma$  upper limits and the extinction-corrected
spectral fluxes of the young  pulsars Crab and PSR B0540$-$69 (Mignani
\& Caraveo 2001), normalized to the N157B distance.  For objects 1, 6,
11 and 12, we plotted only  the 814W spectral fluxes.  We see that the
SEDs for most  of the objects detected in all  the three passbands are
markedly different  from the flattish  spectra of the  young   Crab and PSR B0540$-$69.   This is certainly  the case for
objects 4, 5 and 8, and very likely also for objects 3, 7 and 10, even
when  accounting for  the 
uncertainty  of  the interstellar
extinction correction.  Thus, we  conclude that these objects are most
likely  LMC field  stars, in  agreement with  the results  of  our CMD
analysis,  and  as such  can  be  reasonably  ruled out  as  potential
counterparts to  PSR J0537$-$6910.   The only possible  exceptions are
object 9, which exhibits a  flatter SED than the others, and, perhaps,
object 2. However,  since the latter might actually be  a blend of two
unresolved objects  (see above),  its SED is  probably affected  by an
unknown uncertainty. For objects 1, 6,  11 and 12, the SEDs are poorly
constrained.  However, we  note that for objects 1, 12  and 6 the 555W
and 435W  upper limits indicate rather steep  spectra suggesting that,
whatever  is their  nature, they  can  hardly be  associated with  the
pulsar.  Only for object 11 the constraints on the SED do not allow us
to exclude a flat spectrum.

\begin{figure}[h]
   \centering
      \includegraphics[bb=30 240 420 660,width=8cm,clip]{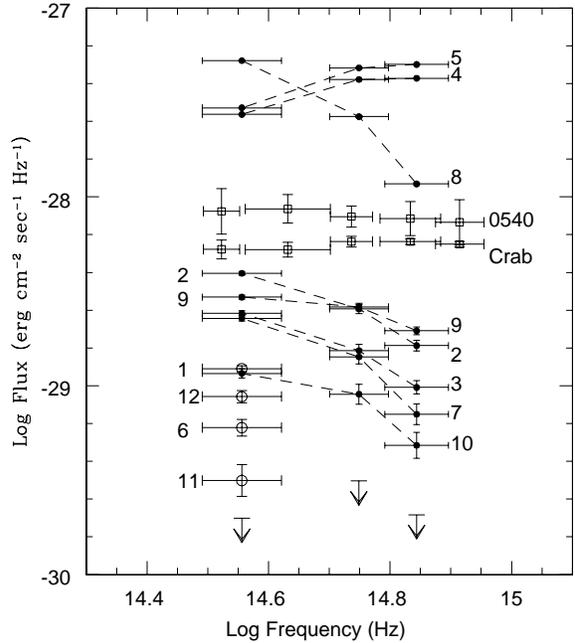}
     \caption{  Spectral   energy   distributions (SEDs) for   the  candidate
counterparts detected in the  three passbands (filled circles) and the
814W spectral  fluxes for the objects  detected in  this passband only
(open circles).   Detection limits for the three filters are 
indicated.  The spectral fluxes were
computed  at  the central wavelengths of  the  ACS passbands and plotted
after    correction  for  an  interstellar   extinction of
$E(B-V)$=0.32,  with $A_{814W}=1.98~ E(B-V)$; $A_{555W}=3.41 ~E(B-V)$ and $A_{435W}=4.12 ~E(B-V)$.  The  numbers correspond to the identifications assigned
in Figure 1.  For comparison  we have plotted the extinction-corrected
SEDs of the Crab and  PSR B0540-69 (empty squares), reproduced  from
Mignani \&  Caraveo (2001), after  normalization to the  N157B distance
(47 kpc).
 } 
\label{spectra} 
\end{figure}

\section{Discussion}

The  results  of  our  multicolor  photometry suggest  that  only  two
objects, 9 and 2, of the eight detected in more than one filter can be
considered as  possible optical  counterparts to the  pulsar, although
object  9 is  among  the most  distant  from the  revised {\sl Chandra}  pulsar
position, and object  2 might be a doublet.  Of  the others, object 11
is  the only  possible candidate.   If either  object 9  or  object 2,
comparable  in  brightness,  were   the  optical  counterpart  of  PSR
J0537$-$6910, it would have an extinction-corrected luminosity $L_{\rm
opt} \approx  9\times 10^{32}$ or $1.2\times  10^{33}$ erg~s$^{-1}$ in
the 814W  passband, for  an assumed pulsar  distance of 47  kpc (Gould 1995).  This
would  imply  that  the  pulsar  has an  optical  emission  efficiency
$\eta_{\rm opt} \equiv  L_{\rm opt}/\dot{E} \approx 1.9\times 10^{-6}$
or  $2.4\times 10^{-6}$.  The  values of  the luminosity  are slightly
below  those of  the young  pulsars Crab  and PSR  B0540$-$69, $L_{\rm
opt}\approx  1.5  \times  10^{33}$  and  $\approx  3  \times  10^{33}$
erg~s$^{-1}$, respectively,  normalized to the 814W  passband.  On the
other hand, the efficiency would  be still comparable with that of the
Crab  ($\eta_{\rm opt}  \approx  3.4 \times  10^{-6}$) but  definitely
below  the efficiency  of PSR  B0540$-$69 ($\eta_{\rm  opt}  \approx 2
\times 10^{-5}$) which is by  far the most efficient optical pulsar to
date (see, e.g., Zharikov et al.\ 2004).

If,  instead, object  11  were  the pulsar  counterpart,  it would  be
fainter  by   a  factor   of  about  10   in  the  optical,   with  an
extinction-corrected luminosity $L_{\rm opt} \approx 9 \times 10^{31}$
erg~s$^{-1}$  and an  efficiency  $\eta_{\rm opt}  \approx 1.9  \times
10^{-7}$.   Finally,  if  none  of  the present  candidates  were  the
counterpart,  the measured  814W upper  limit would  imply  an optical
luminosity  $L_{\rm opt} \le  5.7 \times  10^{31}$ erg~s$^{-1}$  and a
corresponding efficiency $\eta_{\rm opt}  \le 1.2 \times 10^{-7}$.  In
both cases, PSR J0537$-$6910 would be significantly underluminous with
respect to the Crab and PSR B0540$-$69.

 We note that including the uncertainty on the optical extinction
correction affects the above estimates of $L_{\rm opt}$ and $\eta_{\rm
opt}$ by $\le$ 20\% (in the 814W filter) and does not alter significantly our conclusions.

\begin{figure}[h]
   \centering
      \includegraphics[bb=30 240 420 660, width=8cm,clip]{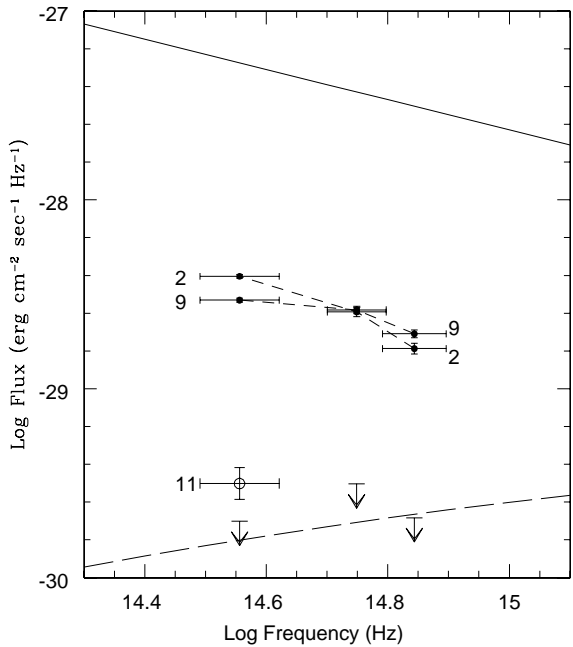}
     \caption{Comparison between the optical SEDs of objects 2, 9 and 11 
 and the extrapolations into the optical domain of the power-law (solid) and 
curved power-law (dashed) models best fitting the {\sl Chandra} 
and {\sl BeppoSAX}/{\sl RXTE} X-ray spectra, respectively (see text). } 
\label{opt_X} 
\end{figure}

It  is  also  interesting  to  confront the  results  of  the  optical
observations with the X-ray data on PSR J0537$-$6910.  We analyzed the
30-ks {\sl  Chandra} ACIS-S3 observation  of the pulsar, taken  in 1/4
subarray  mode  (frame time  0.8  s).   After  correcting for  pile-up
(pile-up fraction was about 5\%--6\%) and subtracting the contribution
from  the  bright PWN,  we  obtained a  power-law  fit  of the  pulsar
spectrum  with a  photon  index $\Gamma  =  1.8\pm 0.1$  and a  photon
spectral  flux  ${\cal  N}   =  (4.4\pm  0.7)\times  10^{-4}$  photons
cm$^{-2}$ s$^{-1}$  keV$^{-1}$ at $E=1$  keV. These values  are almost
independent of the assumed metal abundance in the absorbing (LMC) ISM,
contrary  to  the  hydrogen  column  density: $N_{\rm  H}  =  (0.95\pm
0.07)\times 10^{22}$ cm$^{-2}$ for metal abundances 40\% of solar, and
$N_{\rm  H}  =  (0.50\pm  0.05)\times  10^{22}$  cm$^{-2}$  for  solar
abundances.  The  derived spectral parameters correspond  to the X-ray
luminosity $L_X  = (5.4\pm 1.1)\times 10^{35}$ erg  s$^{-1}$ and X-ray
efficiency  $\eta_X\equiv L_X/\dot{E}  \approx 1.1\times  10^{-3}$, in
the   1--10  keV  band,   significantly  below   those  of   the  Crab
($L_X=1.5\times 10^{36}$ erg s$^{-1}$, $\eta_X=3.4\times 10^{-3}$) and
PSR     B0540$-$69    ($L_X=2.4\times    10^{36}$     erg    s$^{-1}$,
$\eta_X=1.6\times 10^{-2}$; see Zavlin \& Pavlov 2004). \\
While the X-ray and optical efficiencies show very large scatter among
different pulsars, up  to at least 3 orders  of magnitude, their ratio
is  much  more  stable,  typically  $\eta_{\rm  opt}/\eta_X  =  L_{\rm
opt}/L_X \sim 1\times 10^{-3}$ (for the optical and X-ray luminosities
measured  in  the  814W  and  1--10 keV  bands,  respectively),  which
suggests a  common mechanism for the magnetospheric  X-ray and optical
emission  (Zavlin  \&  Pavlov  2004).   If  object 9  (or  2)  is  the
counterpart of PSR J0537$-$6910,  this ratio, $L_{\rm opt}/L_X \approx
1.6\times 10^{-3}$ (or $2.2 \times 10^{-3}$), would be comparable with
the average one for optically detected pulsars. \\
On the  other hand,  if object  11 is the  pulsar counterpart  (or the
counterpart  is  even fainter  than  our  detection  limit), then  the
optical-to-X-ray luminosity ratio  becomes $\approx 1 \times 10^{-4}$,
i.e. the lowest among all pulsars detected in the optical, which
strengthens the case for object 9 or 2 being the pulsar counterpart.\\
For  additional  comparison of  the  optical  and  X-ray data  on  PSR
J0537$-$6910, we show in Figure 4 extrapolations of the pulsar's X-ray
spectrum into  the optical and  the optical spectral fluxes  for three
putative candidates.   We see that the extrapolation  of the power-law
fit into  the optical  exceeds the fluxes  of the  brighter candidates
(objects  9 and  2) by  a factor  of 10--30,  which suggests  that the
broad-band (optical  through X-ray) spectrum cannot be  described by a
simple power law. A similar flattening of the spectrum with decreasing
frequency has  been observed for  other young pulsars (e.g.,  Mineo et
al.\ 2003).  It can be described with a curved power law: $f_E = {\cal
N} E^{-\Gamma(E)}$,  where the  simplest form of  the energy-dependent
photon  index is  $\Gamma(E) =  a +  b\,\log E$  ($f_E$ is  the photon
spectrum, $E$  the photon energy in  keV).  In addition  to the simple
power-law  fit ($b=0$,  $a=\Gamma$),  we  show in  Figure  4 a  curved
power-law fit  ($a=1.33\pm 0.12$, $b=0.15\pm 0.05$)  obtained by Mineo
et  al.\ (2003)  for the  pulsed  X-ray spectrum  of PSR  J0537$-$6910
observed with  {\sl BeppoSAX}  and {\sl RXTE}  in the 1--30  keV band,
where we have  scaled the normalization to the  {\sl Chandra} value to
account for  the unpulsed component.   Extrapolation of this  fit into
the optical  predicts optical fluxes  comparable to the limits  of our
observation, well  below the fluxes of  objects 9 and  2. However, the
example of the  Crab pulsar (Massaro et al.\ 2000)  shows that such an
extrapolation  may underpredict  the  optical fluxes  by  an order  of
magnitude.  Therefore,  the comparison with the X-ray  spectrum of PSR
J0537--6910 suggests that one of  our three candidates is the pulsar's
optical  counterpart,   unless  the  PSR   J0537$-$6910  is  unusually
underluminous in the optical.

\section{Conclusions}

We  have performed deep,  high-resolution, multicolor  observations of
the PSR J0537$-$6910  field with the ACS/WFC on  board {\sl HST}.  Two
likely  candidate  counterparts,  compatible  with  the  revised  {\sl
Chandra} position  of the  pulsar, have been  identified based  on the
color  information.  Although  these candidates  look  interesting for
their spectra, the still limited multicolor information, as well as the
uncertainty of  the interstellar extinction  toward the pulsar,  do not
allow yet  to obtain a  compelling spectral characterization.  For the
third, fainter,  candidate, detected in one filter  only, the spectrum
is virtually unconstrained.  Thus, it is not possible to determine yet
which of the  three, if any, is indeed  the actual pulsar counterpart.
A   follow-up  timing   observation   will  allow   a  more   in-depth
investigation of at least  the two brighter candidates and, hopefully,
will provide  the evidence required  to finally establish  the optical
identification of the pulsar.

\begin{acknowledgements}
 We  are indebted  to  Eric  Gotthelf for  providing  the revised  HRC
 coordinates of the pulsar. It is a pleasure to thanks Guido De Marchi
 for providing us  with the  ACS photometric  calibrations prior to
 publication.   The work  of G.G.P.  was partially  supported  by NASA
 grant NAG5-10865.
\end{acknowledgements}

\end{document}